# Structure refinement from 'Digital' Large Angle Convergent Beam Electron Diffraction Patterns

A. Hubert, R. Roemer and R. Beanland

Department of Physics, University of Warwick, Coventry CV4 7AL

**HIGHLIGHTS**

- We combine of hundreds of convergent beam electron diffraction (CBED) patterns to produce 'digital' large angle CBED (D-LACBED) patterns with angular range > 40 mrad.
- We compare this data with Bloch-wave simulations produced by code running on computing clusters.
- The large amount of data allows parameters to be refined independently and with unique solutions.
- Refinements of atomic coordinates in $Al_2O_3$ have sub-pm precision and accuracy.
- Isotropic Debye-Waller factor (DWF) refinements using an independent atom model (IAM) are precise and accurate for Cu over a range of temperatures.
- Isotropic IAM-DWFs for GaAs are less accurate. Agreement with X-ray data is even worse for $Al_2O_3$.
- Bonding models are necessary for accurate refinements of D-LACBED data.

# Structure refinement from 'Digital' Large Angle Convergent Beam Electron Diffraction Patterns

A. Hubert, R. Römer and R. Beanland

Department of Physics, University of Warwick, Coventry CV4 7AL

**ABSTRACT**: We use semi-automated data acquisition and processing to produce digital large angle CBED (D-LACBED) patterns. We demonstrate refinements of atomic coordinates and isotropic Debye-Waller factors for well-known materials using simulations produced with a neutral, spherical independent atom model. We find that atomic coordinate refinements in $Al_2O_3$ have sub-pm precision and accuracy. Isotropic DWFs are accurate for Cu, a simple fcc metal, but do not agree with X-ray measurements of GaAs or $Al_2O_3$. This indicates that models of bonding will be essential to fully interpret D-LACBED data.

## 1. INTRODUCTION

The unparalleled sensitivity of convergent beam electron diffraction (CBED) to the fine details of electron potential has been known for many years.[1] Despite this strong advantage, it has had limited on studies of electron density and is almost entirely unused for atom bonding and charge density analysis[2, 3]. One of the main reasons for this lack of use is the complexity introduced by multiple scattering, which requires significant modelling and/or simulation, in contrast to the direct access to structure factor amplitudes from methods where single scattering dominates, such as X-ray and neutron diffraction. Another significant barrier to its exploitation is the expertise required to collect good diffraction data from a transmission electron microscope (TEM), and indeed the paucity of data when collected manually. This makes it unattractive in comparison with the well-established and highly automated data acquisition and analysis offered by X-ray techniques. Here, we show that many of these difficulties in the use of electron diffraction can be reduced or eliminated using computer-controlled transmission electron microscopy (TEM) and the vastly increased computing power now widely available for simulations. The combination of hundreds of CBED patterns to produce 'digital' large angle CBED (D-LACBED) patterns has been described previously[4] We compare this data with simulations produced by code running on computing clusters,[5] allowing rapid and robust refinement of crystal potential. To examine the capabilities of the technique we perform refinements of atomic coordinates and isotropic Debye-Waller factors (DWFs) for well-known materials. We find that atomic coordinate refinements in $Al_2O_3$ are impressively accurate, to better than 1pm. Isotropic DWFs are accurate for Cu, a simple fcc metal, but do not agree with the best X-ray measurements of GaAs[6] that take into account bonding and anharmonicity. In $Al_2O_3$, DWF measurements are much larger than accepted values, almost certainly a result of ignoring charge transfer and bonding effects.

Elastic electron scattering is primarily sensitive to the electron potential and on experimental timescales one detects the time-averaged potential, commonly taken to be an incoherent sum over the atomic electron 'cloud' and the thermal vibrations of the atoms. Expressing the potential as a Fourier series allows the convolution of atomic electron density and atom position to be expressed as a product, i.e. using Fourier components of the form

$$V_g = \frac{h^2}{2\pi m_0 e\Omega}\sum_{j=1}^{n} f(s)_j \exp\left(2\pi i(\mathbf{g}\cdot(\mathbf{r}_j + \overline{\mathbf{u}}_j))\right) = \frac{h^2}{2\pi m_0 e\Omega}F_g, \tag{1}$$

where $V_g$ is the Fourier component of the potential for reciprocal lattice vector $\mathbf{g}$, $h$ is Planck's constant, $m_0$ the rest mass and $e$ the charge of the electron, and $\Omega$ the volume of the unit cell. The summation is over the $n$ atoms in the unit cell with mean fractional coordinates $\mathbf{r}_j$ and time-averaged thermal displacements $\overline{\mathbf{u}}_j$. The scattering factor $f_j$ may be viewed as the Fourier transform of the time-averaged position of the electrons in their orbitals around the nucleus of atom $j$. This is a function of the scattering angle, commonly expressed by the parameter $|s| = \sin\theta/\lambda$; if the atom is not spherically symmetric $f_j$ depends upon both the magnitude and direction of $\mathbf{s}$. The sum in Eq. (1) is commonly known as the electron structure factor $F_g$, in analogy with the X-ray structure factor. Nevertheless, for electrons multiple scattering ensures that diffracted intensities are not related in any simple way to the underlying structure factors, unlike X-ray diffraction where $|F_g^2|$ can be obtained directly from the intensity of a diffracted beam.

In an ideal experiment one would measure all Fourier components $V_g$ up to infinite **g**. In reality, of course, only a finite number of **g**-vectors can be sampled. A discontinuity in a Fourier series (such as abrupt truncation) produces oscillatory artefacts in the real space reconstruction of the crystal potential. Unfortunately, these artefacts are quite stubborn and a large number of Fourier components is required to reduce their amplitude, a problem known as the Gibbs phenomenon.[7] This problem can restrict X-ray diffraction,[3] but is even more serious for electron diffraction, which generally is very limited in the number of **g**-vectors that can be accurately sampled.[8] Most refinements based on electron diffraction to date have used a limited number of CBED patterns. The number of **g**-vectors accessible in a single CBED pattern is very limited indeed – only one or two diffracted beams can be set in the Bragg condition and dark field pattern centres[9] (see section 3.1) are rarely accessed. Despite this limitation, precise measurement of individual structure factors was demonstrated by Zuo and Spence in the 1980s[1, 10-14], and it has become common practice to measure a handful of the lowest-order structure factors – which are most sensitive to bonding effects – and use X-ray or neutron diffraction to supply hundreds or thousands of higher order structure factors to complete the picture.[1, 8, 15, 16] However, the combination of disparate data – i.e. from different samples and techniques – may introduce discontinuities in the Fourier series and thus artefacts in real space, which can be of the same magnitude as the bonding effects being sought.[17] Conversely, a theoretical model of the potential is not restricted in the same way and it is straightforward to use functions that are continuous and unbounded. An obvious way to avoid the Gibbs phenomenon is thus to fit experimental measurements to the underlying functions in Eq.1 rather than extracting individual Fourier components, i.e. the diffracted intensities predicted by a model are adjusted to fit experiment through variation of structural parameters.[18] We apply this approach here in a simplistic way, although it has been used with more sophistication and great success in X-ray and neutron diffraction for many years in studies of atomic bonding,[19] by modelling electron shells with pseudopotentials.[2] Currently, there is no equivalent framework for electron diffraction.

An additional complication for electron diffraction is the strong incoherent scattering of electrons channelled down columns of atoms, which requires modelling of electron-phonon interactions to be calculated correctly.[20] This becomes stronger at higher temperatures and electrons scattered in this way contribute to diffuse intensity both outside and under the Bragg scattered beams. Inelastic scattering due to plasmons[21] or Compton scattering[22] can also contribute to diffuse intensity and incoherence. Even worse, electrons scattered in this way may be subsequently Bragg diffracted, producing a strongly structured background that is difficult to subtract. For this reason, energy filtering is often seen as essential for accurate results,[8, 23] which adds yet another experimental complication and barrier to use.

If electron diffraction is to become a routine and widely applied technique that can compete with X-ray and neutron diffraction for structural refinement and bonding studies, both data collection and modelling/refinement must be rapid and straightforward. In our assessment of D-LACBED data here, we therefore compromise on experimental excellence and theoretical rigour in order to obtain measurements quickly and easily. We use a standard workhorse TEM as found in many labs, with no specialist hardware or add-ons such as an energy filter. We find that the results obtained can be sufficiently accurate and precise even without full account of the difficult issues outlined above. This is probably a result of the large amount of data collected.

Simulation of electron diffraction patterns using dynamical diffraction theory can be a time-consuming process. The two current methods in common use, multislice[24] and Bloch wave[25] usually take rather different approaches to the modelling of inelastic scattering. In both implementations, electron-electron scattering is usually neglected. Thermal vibrations are commonly modelled using a frozen phonon approach in multislice calculations, requiring multiple calculations of different configurations; the resulting simulation contains both diffuse and Bragg scattered intensities. Bloch wave simulations usually use the concept of an absorptive (imaginary) potential,[26] i.e. an imaginary component to the scattering factor $if'_j$. attenuating the electron intensity in the Bragg scattered beams. Any subsequent scattering of these 'absorbed' electrons, e.g. back into Bragg scattered beams, is ignored.

Either method could be used to simulate D-LACBED patterns to compare with experimental data. Multislice calculations are suitable for programming with graphical processing unit (GPU) acceleration.[24] Bloch wave calculations require matrix inversion that is often beyond the memory capabilities of a GPU, but can be run efficiently on a cluster of central processing units (CPUs). For our data, with typically $10^6 - 10^7$ pixels in a single dataset, it is not immediately apparent which method gives the best combination of speed and accuracy. In keeping with most previous CBED studies we calculate diffracted intensities with the Bloch wave method. We use message passing interface (MPI) parallelisation, typically using 200-400 cores to obtain high quality simulations in tens of seconds.

Revisiting equation (1) with the addition of an absorptive potential we have

$$V_g = \frac{h^2}{2\pi m_0 e\Omega}\sum_{j=1}^n \big(f(\mathbf{s})_j + if'(\mathbf{s},\overline{\mathbf{u}})_j\big)\exp\big(2\pi i(\mathbf{g}\cdot(\mathbf{r}_j + \overline{\mathbf{u}}_j)\big), \tag{2}$$

where $f'_j$ is generally a function both of $\mathbf{s}$ and $\overline{\mathbf{u}}_j$. Evaluation of Eq. (2) in the general case is not straightforward. Considerable simplifications can be made with the assumption of spherical atoms, in which case a scalar $s$ may be used, while the Einstein model of independent, harmonic thermal vibrations, allows $\overline{\mathbf{u}}_j$ to be converted to an isotropic temperature factor $\exp(-B_j s^2)$, (i.e. the Fourier transform of the time-averaged position of the atom, as it moves about its mean position due to thermal vibrations) where the Debye-Waller factor $B = \overline{u}_j^2/16\pi^2$ is determined by the mean square thermal displacements $\overline{u}_j^2$, i.e.

$$V_g = \frac{h^2}{2\pi m_0 e\Omega}\sum_{j=1}^n \big(f(s)_j + if'(s,B)_j\big)\exp\big(2\pi i(\mathbf{g}\cdot\mathbf{r}_j)\exp(-B_j s^2). \tag{3}$$

In this simplified framework, the absorptive potential $f'_j$ due to thermal diffuse scattering may be calculated from a knowledge of $f(s)_j$ and $B_j$ with no free parameters using e.g. the Bird and King model.[26] If we use calculated scattering factors $f(s)_j$ the remaining parameters accessible to experiment are simply the coordinates $\mathbf{r}_j$ and Debye-Waller factor for each atom in the unit cell. Here we test the ability of D-LACBED to measure these parameters for simple well-known materials.

## 2. EXPERIMENTAL AND COMPUTATIONAL METHODS

Electron transparent specimens were prepared using standard methods, i.e. mechanical grinding and polishing followed by $Ar^+$ ion milling. Surface damage was minimised for the GaAs and $Al_2O_3$ specimens with a final low energy (2keV) mill for 15 minutes, while the Cu specimen was etched briefly in 1% Nital solution. Data were collected using a JEOL 2100 LaB$_6$ TEM operating at 200 kV equipped with a Gatan Orius SC600 camera. Prior to data collection, Digital Micrograph™ scripts were used to calibrate both beam displacement and image shift caused by spherical aberration of the pre-field and post-field objective lenses respectively. Data acquisition was performed with a third script that collected CBED patterns for many incident beam orientations, using the previous calibrations to apply compensating beam shifts to maintain the position of the electron beam on the specimen to an accuracy of ~1nm while doing so. Roughly ten CBED patterns were captured per second, each with a size of 672x668 pixels. The electron beam typically had a convergence angle >1 mrad focused to a probe of ~8nm FWHM on the specimen. Heating experiments were performed using a Gatan 952 double-tilt heating holder.

Processing of the acquired CBED data to obtain D-LACBED patterns suitable for comparison with simulation was performed as follows. Since we generally use relatively small convergence angles, the CBED discs are almost featureless and we simply extrapolate diffuse intensity from outside the discs to subtract the background. A 2D cubic spline was fit to the diffuse scattering between the discs and subtracted from each CBED pattern before using simple cut and paste, averaging where beams overlap, to assemble a set (typically between 20 and 200) of D-LACBED patterns.

Linear distortions in the patterns due to residual intermediate lens astigmatism and other lens aberrations were corrected using the known symmetry of the 000 pattern and/or comparison with simulations to measure any skew and stretch (typically <0.5%). With the large angles and small camera lengths used here (typically 200mm) higher-order Laue zone (HOLZ) lines have sub-pixel widths and are essentially invisible. Thus, the zero-order Laue zone (ZOLZ) symmetries of the collection of LACBED patterns were used to reduce noise by averaging. Finally, the patterns were rotated and cropped to be square, typically between 250 and 400 pixels, giving a useful experimental dataset of ~2x10^6 to ~3x10^7 measured intensities. The point spread function of the camera has a noticeable effect on the data. This (and any blurring of the data due to a lack of energy filtering) was accounted for by applying a Gaussian blur (typical radius 2.2 pixels) to the simulation in preference to application of a deconvolution to the experimental data.

Simulations were performed using standard Bloch-wave methods[24, 25] parallelised to run on a computing cluster.[5, 27, 28] For each pixel, the Bloch waves ('strong beams') used in the calculation were chosen according to their perturbation strength as defined by Zuo and Wieckenmeier.[29] It was found that perturbative inclusion of 'weak beams', while much quicker to calculate, had almost negligible effects on the calculation and did not compensate for missing strong beams at the required accuracy. Absorptive potentials were calculated using the Bird and King method[26] with isotropic Debye-Waller factors and Kirkland scattering factors.[24] The simulation time $T$ scales as the cube of the number of Bloch waves $N$ included in the calculation. Timings were optimised by analysing differences in intensity for simulations with different $N$,[30] typically giving $100 < N < 400$ for different datasets and $T < 2$ minutes using up to 280 cores for unbinned data and a single specimen thickness.

Zero-mean normalised cross-correlation[31] (ZNCC) was used to quantify the fit between experimental and simulated D-LACBED patterns. This has the advantage of being insensitive to background offsets and scaling, at the expense of sensitivity to lateral displacements. Thus, a small sub-pixel shift of the experimental data was generally required to obtain the best possible fit to simulation. Fits were obtained for each D-LACBED pattern individually, removing any dependence on relative intensities of different reflections. To give some resemblance to the $R$-factor commonly used to indicate experiment/simulation fit quality in structure solution methods the fit index for $n$ experimental patterns $x$ and simulated patterns $y$, each with $N$ pixels is

$$f = \sum_{j=1}^{n} \left( 1 - \sum_{i=1}^{N} \frac{(x_{i,j} - \bar{x}_j)(y_{i,j} - \bar{y}_j)}{\sigma_{xj}\sigma_{yj}N} \right)_j \tag{4}$$

where the first sum is over all patterns and the second sum is the ZNCC: $\bar{x}_j$ and $\bar{y}_j$ are the means of the $j^{th}$ experimental and simulated patterns respectively; $x_{i,j}$ and $y_{i,j}$ are their pixel values; $\sigma_{xj}$ and $\sigma_{yj}$ are their standard deviations. Equation (4) gives $f = 0$ for a perfect fit.

## 3. RESULTS

### 3.1 Copper.
The simplest possible test of the capabilities of D-LACBED would be a material that has a limited number of parameters that can be refined and conforms reasonably well to the assumptions of the theoretical model (i.e. spherical, neutral atoms that are well-described by calculated scattering factors, with harmonic thermal vibrations). Most monatomic metals fit this description and we choose copper here due to its ready availability. The atomic coordinates of this fcc crystal are fixed, leaving only Debye-Waller factor as a measurable parameter. We collected data from [001] and [114] zone axes as examples of relatively dense low-index diffraction patterns and more sparse mid-index patterns respectively. The small lattice parameter of Cu results in relatively large Bragg angles (e.g. $2\vartheta_{002}$ = 13.8 mrad), allowing large convergence angles without overlapping discs in the CBED pattern. We thus used a beam half-convergence angle of 5.9 mrad, requiring only ~120 CBED patterns to produce D-LACBED data extending beyond 40 mrad. Figure 1 shows some of the D-LACBED patterns collected from Cu at room temperature (303 K) from the [001] and [114] zone axes. As defined by Buxton et al.[9], the centre of a dark field LACBED pattern is marked by a point of symmetry, which must be at least a 2-fold axis for diffraction in the zero-order Laue zone (ZOLZ). The symmetry may be higher; for example, the 220 and 440 patterns in Fig. 1 have ZOLZ symmetry mm2 and the pattern centre is marked by a bright point for both. Note that – since all patterns sample the same field of view, dictated by the camera and crystal orientation – all dark field pattern centres are displaced from the centre of the field of view towards the 000 pattern. The [001] zone axis has symmetry 4*mm*, meaning that although 25 complete D-LACBED patterns were collected there are only six unique types, i.e. 000, 200, 400, 220, 420 and 440. The [114] ZOLZ symmetry is *mm*2, giving five unique patterns from the thirteen displayed in Fig. 1b plus the 442-type reflection (not shown). Data similar to Fig. 1 was collected at 373K and then at temperature increments of 100K to a maximum of 753K.

It was found that optimised simulations required >210 Bloch waves for [001], but only >120 for [114] data.[30] The smaller number for [114] reflects the larger **g**-vectors present in the data, giving a more sparsely populated ZOLZ section through the reciprocal lattice and correspondingly fewer Bloch waves required. The 400x400 pixel simulations at a single thickness required 1 minute 10 sec for Cu [001] and 50 seconds for Cu [114] using 256 cores. The first step to compare with simulation is to obtain a good fit for the specimen thickness. Although this is a basic parameter required for any measurement, it can be problematic for conventional CBED due to the small size of the discs and limited data. This usually results in strong oscillations of fit parameter with specimen thickness[32, 33]. Conversely, the large amount of LACBED data gives a single, well-defined thickness that gives the best fit to the data (inset Fig. 1a).

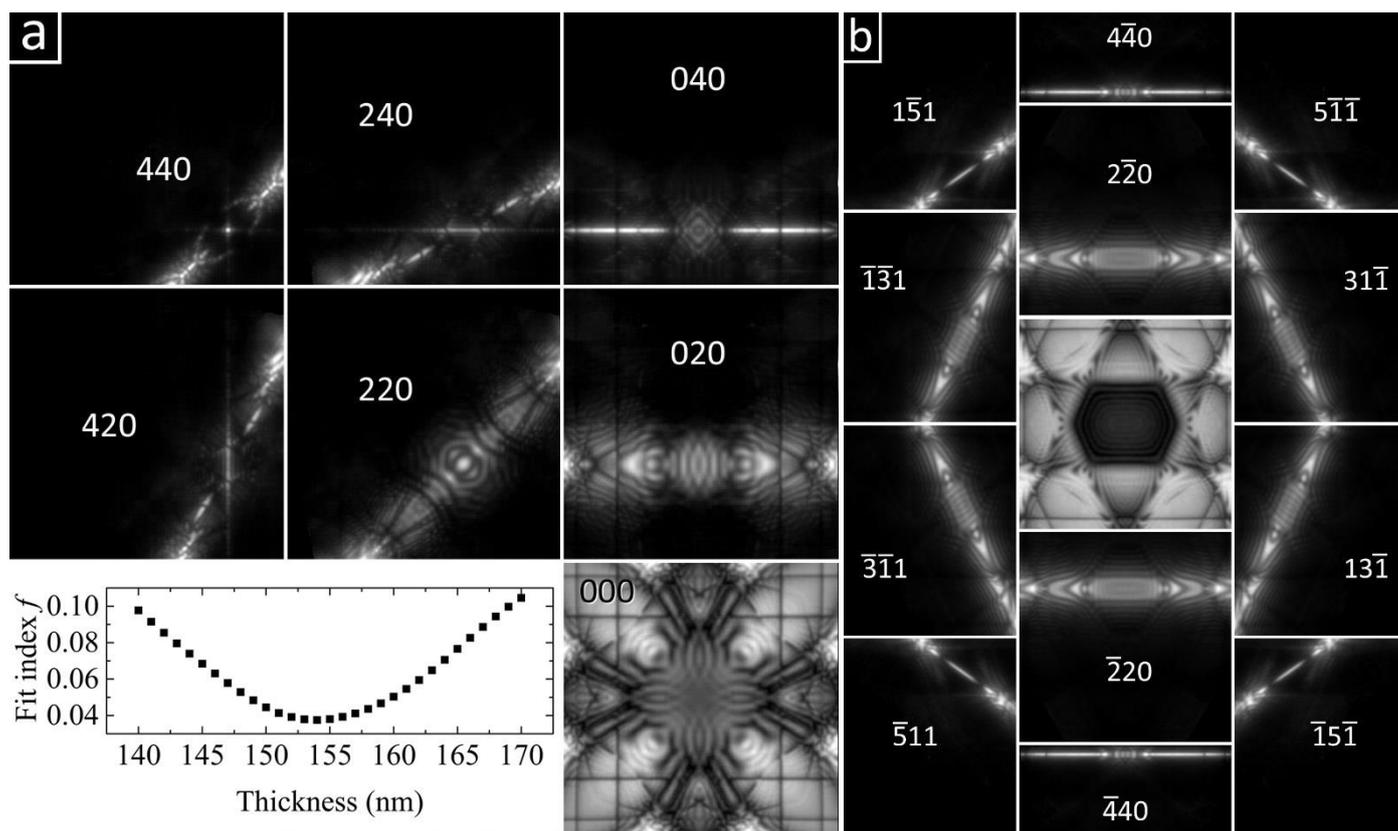

Figure 1. Room temperature D-LACBED from Cu (a) Seven patterns from Cu [001], angular range 41.3 mrad. Inset shows the fit index for simulated patterns as a function of specimen thickness. (b) Thirteen patterns from Cu [114], angular range 44.2 mrad. All patterns are normalised for display here to the visible display range and have applied gamma 1.5 to allow features in darker parts of the image to be seen more easily.

The fit to simulation was refined by varying the isotropic Debye-Waller factor $B$. The plots in Fig. 2 shows the fit index (Eq. 4) for the different temperatures as a function of $B$. This demonstrates a well-behaved parameter space with a single minimum; fit indices were typically below 5% but increased slightly for higher temperatures. Simple gradient-descent optimisation is sufficient to determine the best fit to experiment, typically requiring less than ten iterations to obtain a precision of 0.01 Å². Again, this is a significant improvement over CBED data, which often show many local minima[34]. Experimental 000 patterns and best fit simulations for the Cu [001] data are also shown in Fig. 2. The most obvious change with increasing temperature is the darkening of the bands running along <100> directions through the centre of the pattern. This is due to increased thermal scattering, which strongly affects the absorptive potential and gives increased absorption in orientations where strong channelling occurs. A subtler effect is also present, i.e. the weakening of sharp features such as the horizontal and vertical 400-type deficit lines with increasing temperature.

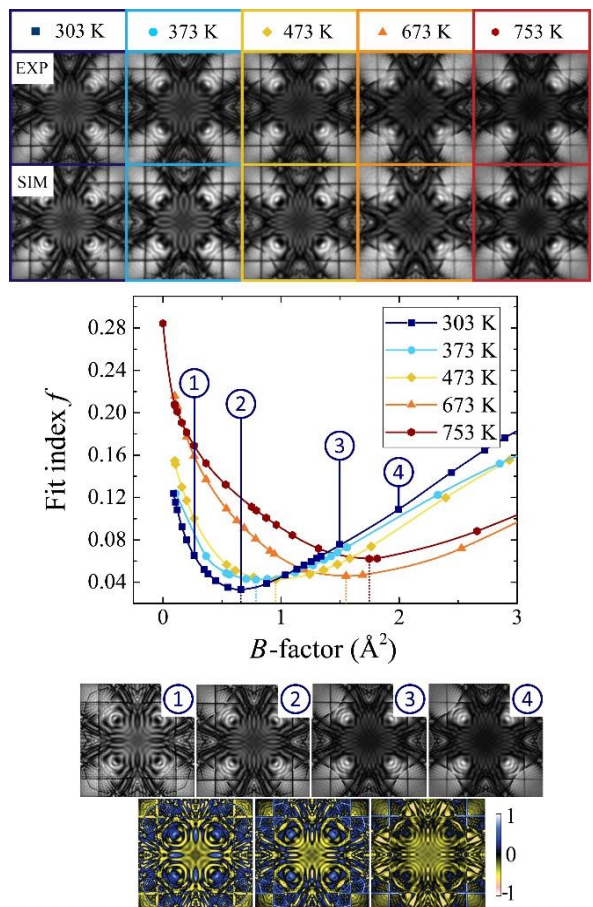

Figure 2. Upper: the experimental 000 D-LACBED patterns and best fit simulations at different temperatures for Cu [001]. Centre: fit indices $f$ (all patterns) for a range of simulations at each temperature. Lower: simulated 000 patterns (1) – (4) and the difference between them, where yellow shows a decrease, and blue an increase, in relative intensity with increasing Debye-Waller factor $B$.

Simulated 000 patterns at fixed specimen thickness with $B$ (1) to (4) are also shown in Fig. 2. The difference images beneath show the changes that affect the fit index $f$, i.e. variations in *relative* intensity when each pattern is normalised to the same range, rather than changes in absolute intensity. This is useful here since it corresponds to the normalised cross-correlation used to calculate the fit index (Eq. 4). These complicated fringe patterns exhibit both increases and decreases in relative intensity. Additionally, the 400-type deficit lines show a relative decrease in intensity at low $B$ that becomes a relative increase at higher $B$, showing an interplay between absorption and thermal effects. The initial darkening of the line is caused by absorptive scattering, which affects large **g**-vectors more readily than small ones at small values of $B$.[26] At higher temperatures however, these diffraction features become continuously weaker (brighter) as expected from Eq. (3). Cu [411] data shows similar behaviour (Fig. S2). Here the central dark lozenge is most affected by absorption and again fine linear features become less visible at higher temperatures.

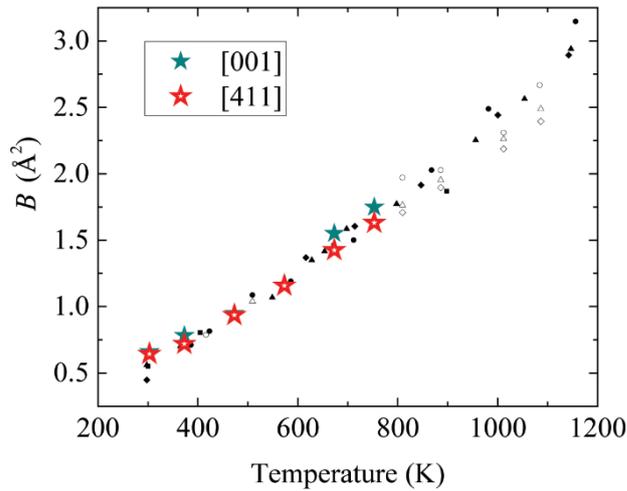

Figure 3. Experimental determinations of Debye-Waller factor $B$ in copper D-LACBED data from [001] (solid stars, green) and [114] (hollow stars, red). Errors are smaller than the data points, typically ~0.01 Å². Many previous measurements of $B$ using X-ray diffraction and the Mössbauer effect are shown in black and white (from ref [35])

The measured Debye-Waller factors for copper as a function of temperature are shown in Fig. 3, together with historical data collated by Shukler [35]. Excellent agreement is found, showing that D-LACBED data gives accurate temperature factors in this simple metal. Nevertheless, there is a small disagreement between measurements from the [001] and the [114] measurement, particularly at higher temperatures. It is well-known that the deviation of the data in Fig. 3 from a straight line is mainly due to anharmonic thermal vibrations, which become more significant at higher temperatures.[36] This cannot be captured by the single Debye-Waller factor, which only describes harmonic vibrations; attempts to fit experimental data from an anharmonic material using a Debye-Waller factor will effectively give slightly differing answers for different **g**-vectors. This can be used to advantage in measurements using the Mössbauer effect,[36] where comparison of Debye-Waller factors obtained with first and second order diffraction can be used to calculate anharmonic components. Such an approach is unlikely to be successful in the case of D-LACBED data, since dynamical diffraction mixes intensities between the different reflections (for example, in Fig. 1a the transfer of intensity from the 040 to the 420 and 440 patterns is quite clear). Here, the Debye-Waller factors are derived from a fit to all the D-LACBED patterns at a zone axis and are thus some kind of average measurement. Nevertheless, since the patterns all lie in the ZOLZ, the [001] and [114] D-LACBED data cover different parts of reciprocal space and it is perhaps not surprising that small differences in measured Debye-Waller factors appear when anharmonic thermal vibrations are known to be present.

### 3.2 Sapphire

Perhaps the most useful application of D-LACBED may be to refine atomic coordinates in materials that contain a large fraction of elements with low atomic number (which scatter electrons more strongly than X-rays[37]) or that are only available in nanoscale form, so that X-ray diffraction becomes difficult due to Scherrer broadening. Surprisingly, this does not appear to have been attempted with CBED refinements. A test material for this application must have some atomic coordinates that are not completely fixed by the crystal symmetry; here, we choose α-Al₂O₃, (space group $R3c$). In the hexagonal setting the Al atoms in the two-atom basis have coordinates [0, 0, $z_{Al}$], while O atoms have fractional coordinates [$x_O$, 0, ¼], i.e. each atom has a single coordinate not fixed by the space group that can be refined experimentally. High-quality X-ray measurement[38] gives $z_{Al}$ = 0.352156(17) and $x_O$ = 0.69364(7) at room temperature.

D-LACBED data were collected from the [2$\overline{2}$1] zone axis, which has a smallest Bragg angle of $2\vartheta_{012}$ = 7.2 mrad at 200 kV. 961 CBED patterns were collected with a beam half-convergence of 1.7 mrad, giving D-LACBED data extending beyond 40 mrad. The reconstructed D-LACBED patterns had dimensions of 296 x 296 pixels. The ZOLZ has symmetry $mm2$, allowing the unique patterns to be displayed in one quadrant. Figure 4 (top left) shows 16 unique patterns taken at room temperature (29 °C). Refinement of atomic coordinates from this data gave $z_{Al}$ = 0.35246(5) and $x_O$ = 0.6932(1), i.e. a difference of 0.4 pm and 0.2 pm respectively from the X-ray values.[30] As shown in Fig. 4 bottom right, the data gives a unique best fit that is easily found using two-dimensional gradient descent. To illustrate the sensitivity of D-LACBED data to atomic coordinates we show the changes in the patterns using a form of normalised differential (Fig.4 top right and bottom left) that is appropriate for our use of ZNCC as a fit index. (Intensity differences are calculated between simulations for small changes $\delta x_O$ and $\delta z_{Al}$. Each pattern's mean is subtracted and intensity normalised to give a standard deviation of unity. The scales of $\delta I/\delta x_O$ and $\delta I/\delta z_{Al}$ in Fig. 4 are thus in units of standard deviation.) It is clear from these images that all patterns display strong sensitivity to sub-picometre atomic displacements of both atoms. Some regions in some patterns are very strongly affected with normalised differentials of almost 100 (often close to, or at, the pattern centre), while other regions are relatively insensitive. Importantly, the intricate changes intensity are quite different for the two parameters, which results in the ability to refine them essentially independent of each other.

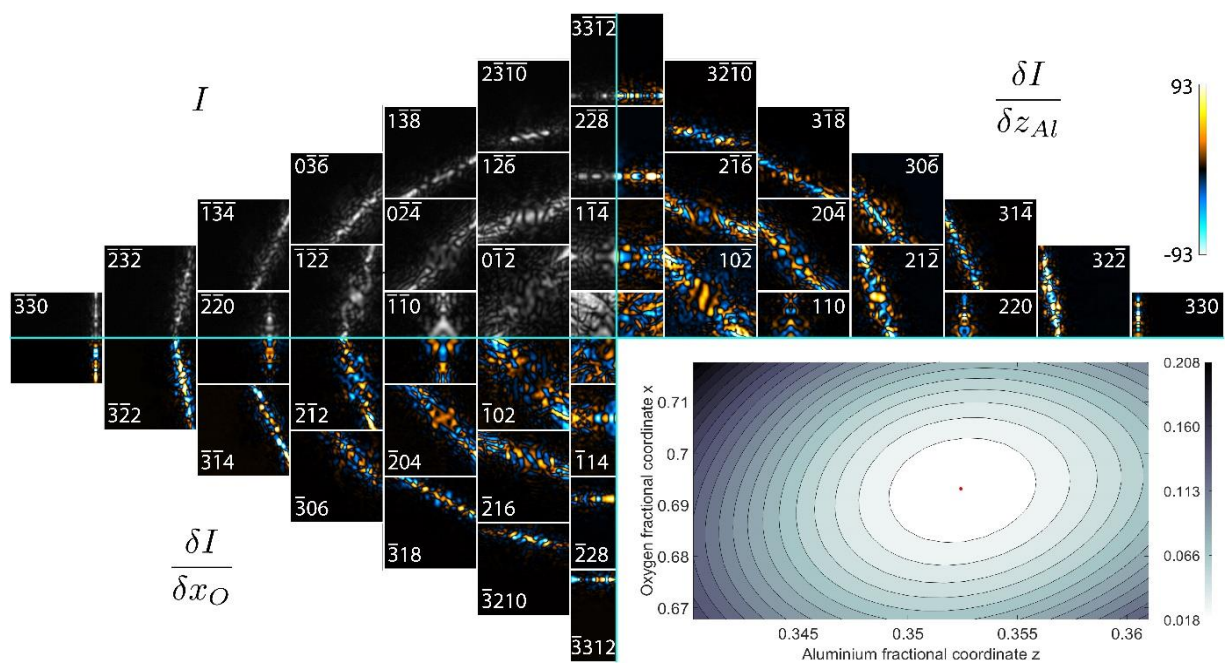

Figure 4. Top left: D-LACBED data from $[2\bar{2}1]$ $\alpha$-Al₂O₃, each pattern has an angular range 40.6 mrad. Bottom right: fit index $f$ as a function of $x_O$ and $z_{Al}$.. Top right: $\delta I/\delta z_{Al}$, and bottom left: $\delta I/\delta x_O$.

### 3.3 Gallium arsenide

While the above examples indicate that accurate measurements are possible from D-LACBED data, it was found that refinement of DWFs for sapphire gave values of $B_{Al}$ = 0.15 and $B_O$ = 0.44 Å², much larger than the X-ray values of $B_{Al}$ = 0.00327(6) and $B_O$ = 0.00387(6) Å² respectively.[38] These gross differences are probably a result of bonding and charge transfer effects, i.e. the failure of the neutral, independent atom model for Al³⁺ and O²⁻. We thus examine a material that has smaller, but still significant, bonding effects without the complication of any atomic coordinate refinement, i.e. GaAs. The smallest Bragg angle in the GaAs $[\bar{1}10]$ pattern is $2\vartheta_{111}$ = 7.7 mrad at 200 kV, requiring much smaller convergence angles than Cu to avoid overlapping discs in the CBED pattern. We used a beam half-convergence angle of 1.16 mrad and 1681 CBED patterns to produce D-LACBED data extending beyond 40 mrad. The reconstructed D-LACBED patterns had dimensions of 320x320 pixels. Figure 4 shows eighty-five patterns taken at room temperature (29 °C) and 200 °C. The vertical (110) mirror symmetry reduces the number of unique patterns to forty-nine.

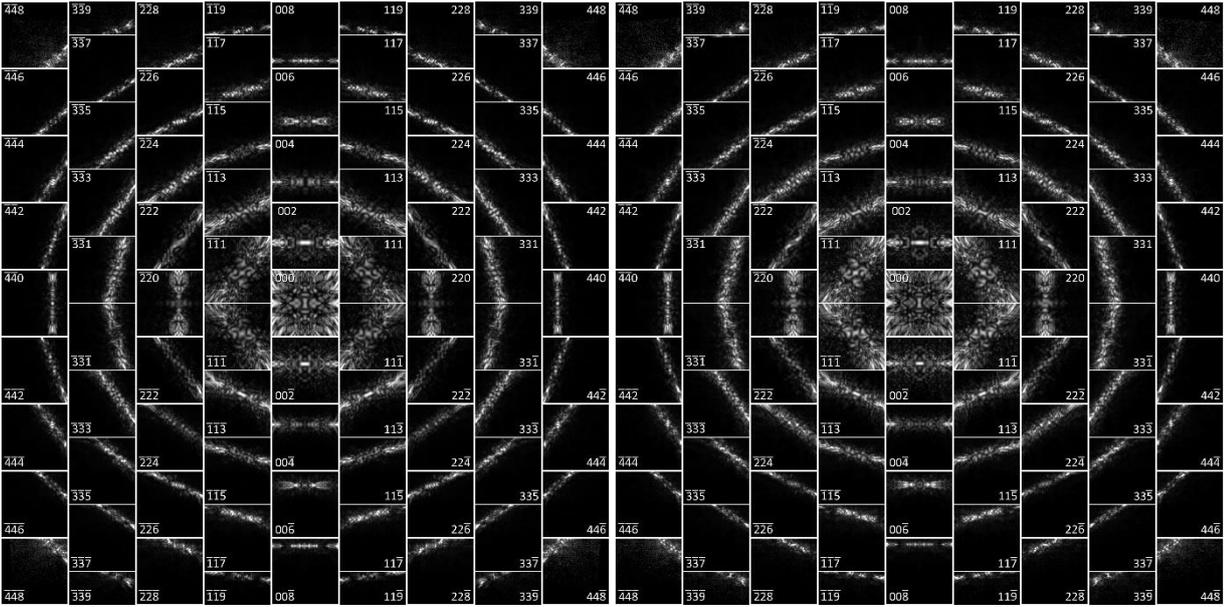

Figure 5. Eighty-five D-LACBED patterns from [[1$\bar{1}$0]] GaAs. Left: room temperature (29 °C). Right: 200 °C. Each pattern has an angular width of 46.2 mrad.

Optimising the fit between simulation and experiment for $B_{Ga}$ and $B_{As}$ using two-dimensional gradient descent gives a unique solution at $B_{Ga}$ = 0.83(2), $B_{As}$ = 0.68(2) Å$^2$ for the room temperature data with a fit $f$ = 4.61% and $B_{Ga}$ = 1.01(2), $B_{As}$ = 0.80(2) Å$^2$ ($f$ = 4.96%) at 200 °C. The normalised differentials $\delta I/\delta B_{Ga}$ and $\delta I/\delta B_{As}$ for the central few D-LACBED patterns are shown in Figs. 6d and e (see also [30]). The influence of DWFs here on the ZNCC, while significant, is roughly two orders of magnitude smaller than the effect of atomic coordinates in the sapphire refinement. As observed for the above refinements, changes in the parameters $B_{Ga}$ and $B_{As}$ produce a complicated pattern of increases and decreases in intensity resulting in an effective independence of the fits shown in Figs. 6a and b. Typically, fewer than twenty iterations were required to find the best fit to a precision better than 0.01 Å$^2$. While the time to obtain a result is acceptable for this simple two-parameter problem, it is still quite long if more complex problems are to be tackled. Thus, we tested the reproducibility of the result using data binned by 2 (160x160 pixels) and binned by 4 (80x80 pixels), giving simulation times of ~22 seconds and ~12 seconds respectively using the same simulation conditions. These gave $B_{Ga}$ = 1.04, $B_{As}$ = 0.78 Å$^2$ and $B_{Ga}$ = 1.11, $B_{As}$ = 0.88 Å$^2$ respectively, i.e. an error of up to 10% for a reduction in time of roughly eight times.

Comparison of these results with literature values (Fig. 5c) shows considerably less agreement than was obtained in the case of Cu. The most accurate measurement of Debye-Waller factors in GaAs using X-ray diffraction gives $B_{Ga}$ = 0.622(3), $B_{As}$ = 0.483(5) Å$^2$ at room temperature.[6] It was noted in this X-ray study that both bonding effects (for small $\mathbf{g}$-vectors) and anharmonicity (for large $\mathbf{g}$-vectors) must be taken into account for an accurate measurement. Since we do not consider either of these effects here, it is perhaps not surprising that our measurement does not agree. Interestingly, all experimental measurements are clustered along a single trend line and our measurement is fairly close to another measurement made by electron diffraction.[23] Fig. 6f shows the difference between best-fit simulation and experiment at room temperature. We take the systematic nature of the residual intensity to indicate that refinement against parameters such as bonding and/or anharmonicity would produce a better fit.

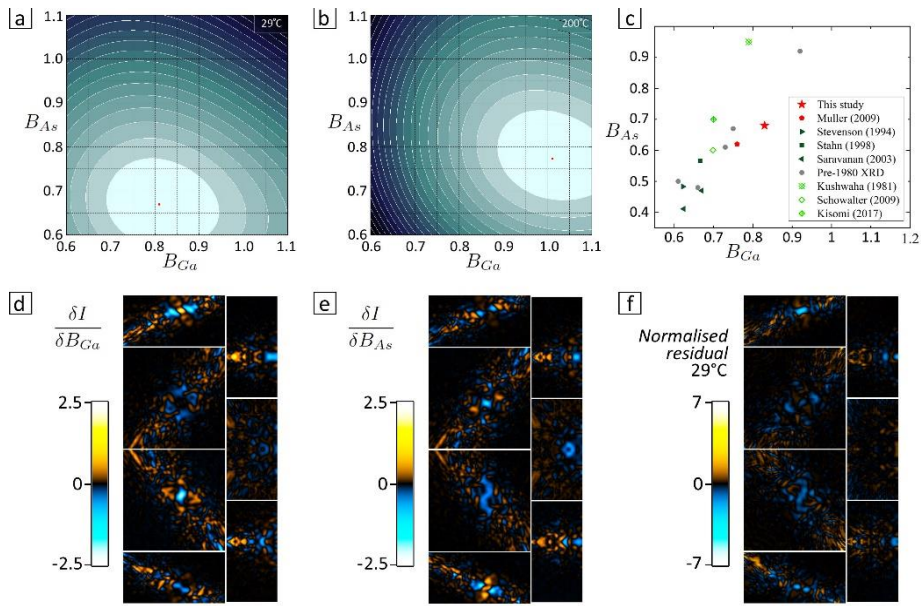

Figure 6. a, b: fit index $f$ (49 unique patterns) for [110] GaAs data at room temperature and 200°C as a function of $B_{Ga}$ and $B_{As}$. c: Experimental measurements of $B_{Ga}$ and $B_{As}$. Green and gray = X-ray diffraction[6, 39-45]; red = electron diffraction[23]; green = theory[46-49]. Unattributed measurements are those collated in ref. [6]. d,e: $\delta I/\delta B_{Ga}$ and $\delta I/\delta B_{As}$ in units of standard deviation. f: the residual between best-fit simulation and experiment, calculated in the same way as d and e. (See also [30]).

## 4. DISCUSSION

In principle, the interference effects giving rise to the complicated patterns seen in dynamical diffraction patterns can be used to obtain both the amplitude and phase of every diffracted beam[50] and much of the literature concerns the extraction of phase relationships by considering e.g. 3-beam diffraction,[28, 51, 52] or points with specific symmetries.[53] We do not consider such details here. The main point is that these studies show that there are specific regions in D-LACBED patterns that have intensities which are very strongly dependent upon the amplitude and phase of structure factors. This, together with a large amount of data, is sufficient for a 'brute force' minimisation approach to obtaining the best match between simulation and experiment. Nevertheless, it is clear that the multitude of special points in D-LACBED data[27] could be further exploited with a more intelligent approach.

Often, the goal of CBED studies has been to obtain information about $f_j$, in particular structure factors that indicate deviations from spherical atomic symmetry.[8, 10, 16, 54] The examples here show that there is much to be gained by examining the other parameters that influence dynamical electron diffraction patterns. In particular, the ability to refine atomic coordinates to sub-picometre precision using nanoscale probes will have applications in many fields. Fig. 4 suggests that accuracy is a picometre or better, even when bonding is neglected.

Since both electron orbitals and thermal vibrations are averaged over the timescale of the measurement, it is not generally possible to separate their influence on the data. The presence of strong bonding effects is probably the reason for the incorrect measurement of Debye-Waller factors in $Al_2O_3$ and GaAs here. There are obvious opportunities to apply multipole modelling[2, 3] or Hirshfield refinement[55] methods to electron data.

Finally, it should be mentioned that almost every aspect of the work presented here can be improved. Control of the TEM through scripts on a third-party computer is relatively slow; we estimate that a hardware-synchronised system using a fast camera could collect data, similar to that shown here, in a second or two. The angular range of the data can be optimised; it is not clear at present how many (or how few) data is needed to allow an accurate refinement. Since all the data is essentially due to ZOLZ diffraction (although HOLZ lines can be seen in D-LACBED data of higher quality),[4] it is insensitive to atomic displacements along the electron beam direction. While this might restrict the choice of zone axis in specific cases, more often the presence of symmetrically-equivalent atoms ensures that a component parallel to the beam for one atom is a perpendicular component for another. The zone axes used for the investigations here were dictated primarily by their ease of access in the specimens examined. Our simulations, although sufficiently rapid and accurate as deployed on a powerful cluster, do not include all aspects of the electron specimen interaction. Gradient descent may not be the most efficient algorithm to find the best fit to experiment for D-LACBED data and indeed there may be more sensitive ways to measure large changes in intensity in specific parts of a large dataset than ZNCC. Nevertheless, despite these deficiencies, it is clear that D-LACBED data can yield useful and interesting results that cannot be obtained with other scattering methods.

## 5. CONCLUSIONS

We have investigated the suitability of D-LACBED for structure refinement within the constraints of an independent atom model (IAM). We find that the large amount of data generally results in a unique best fit between experiment and simulation that is easily found by simple gradient descent. Refinement of Debye-Waller factors for a material that is well-described by the IAM, (i.e. copper) agrees with other measurements. This is not the case for $Al_2O_3$ or GaAs, almost certainly due to the lack of bonding and anharmonicity in our model. Refinement of atomic coordinates for $Al_2O_3$ is obtained, agreeing with literature values with sub-pm precision and accuracy.

### ACKNOWLEDGEMENTS

AH would like to acknowledge EPSRC funding EP/M506679/1 and EP/M508184/1.

### DATA REPOSITORY

Additional information and raw data can be found at http://wrap.warwick.ac.uk/109737.

1. Spence, J., *On the accurate measurement of structure-factor amplitudes and phases by electron diffraction.* Acta Crystallographica Section A, 1993. **49**(2): p. 231-260.
2. Coppens, P., *X-ray charge density analysis and chemical bonding*. 1 ed. IUCr Texts on Crystallography. 1997: Oxford, UK. 358.
3. Gatti, C. and P. Macchi, *Modern Charge Density Analysis*. 2012, Dordrecht: Springer.
4. Beanland, R., et al., *Digital electron diffraction - seeing the whole picture.* Acta Crystallographica Section A, 2013. **69**(4): p. 427-434.
5. *FELIX Bloch wave simulation: Source code* Available from: https://github.com/RudoRoemer/Felix.
6. Stevenson, A.W., *Thermal Vibrations and Bonding in GaAs - an Extended-Face Crystal Study.* Acta Crystallographica Section A, 1994. **50**: p. 621-632.
7. Hewitt, E. and R.E. Hewitt, *The Gibbs-Wilbraham phenomenon: An episode in fourier analysis.* Archive for History of Exact Sciences, 1979. **21**(2): p. 129-160.
8. Nakashima, P.N.H., et al., *The Bonding Electron Density in Aluminum.* Science, 2011. **331**(6024): p. 1583-1586.
9. Buxton, B.F., et al., *The Symmetry of Electron Diffraction Zone Axis Patterns.* Philosophical Transactions of the Royal Society of London. Series A, Mathematical and Physical Sciences, 1976. **281**(1301): p. 171-194.
10. Zuo, J.M., J.C.H. Spence, and M. Okeeffe, *Bonding in GaAs.* Physical Review Letters, 1988. **61**(3): p. 353-356.
11. Spence, J.C.H., J.M. Zuo, and R. Hoier, *Accurate Structure-Factor Phase Determination by Electron-Diffraction in Noncentrosymmetric Crystals - Reply.* Physical Review Letters, 1989. **63**(10): p. 1119-1119.
12. Zuo, J.M. and J.C.H. Spence, *Automated Structure Factor Refinement from Convergent-Beam Patterns.* Ultramicroscopy, 1991. **35**(3-4): p. 185-196.
13. Zuo, J.M., *Automated Structure-Factor Refinement from Convergent-Beam Electron-Diffraction Patterns.* Acta Crystallographica Section A, 1993. **49**: p. 429-435.
14. Zuo, J.M. and J.C.H. Spence, *Experimental test of accuracy in automated structure refinement from CBED*. Electron Microscopy 1994, Vol 1: Interdisciplinary Developments and Tools, ed. B. Jouffrey and C. Colliex. 1994. 851-852.

15. Zuo, J.M., et al., *Quantitative electron diffraction and applications to materials science*. Microbeam Analysis 1995: Proceedings of the 29th Annual Conference of the Microbeam Analysis Society, ed. E.S. Etz. 1995. 269-270.

16. Zuo, J.M., et al., *Direct observation of d-orbital holes and Cu-Cu bonding in Cu2O.* Nature, 1999. **401**(6748): p. 49-52.

17. Bernard, J.E. and A. Zunger, *Bonding charge density in GaAs.* Physical Review Letters, 1989. **62**(19): p. 2328-2328.

18. Koritsanszky, T.S. and P. Coppens, *Chemical applications of X-ray charge-density analysis.* Chemical Reviews, 2001. **101**(6): p. 1583-1627.

19. Stewart, R., *Electron population analysis with rigid pseudoatoms.* Acta Crystallographica Section A, 1976. **32**(4): p. 565-574.

20. Martin, A.V., S.D. Findlay, and L.J. Allen, *Model of phonon excitation by fast electrons in a crystal with correlated atomic motion.* Physical Review B, 2009. **80**(2): p. 024308.

21. Allen, L.J., A.J. D'Alfonso, and S.D. Findlay, *Modelling the inelastic scattering of fast electrons.* Ultramicroscopy, 2015. **151**: p. 11-22.

22. Eaglesham, D.J. and S.D. Berger, *Energy filtering the "thermal diffuse" background in electron diffraction.* Ultramicroscopy, 1994. **53**(4): p. 319-324.

23. Muller, K., et al., *Refinement of the 200 structure factor for GaAs using parallel and convergent beam electron nanodiffraction data.* Ultramicroscopy, 2009. **109**(7): p. 802-814.

24. J., K.E., *Advanced computing in electron microscopy*. 2010, New York: Springer

25. A.J.F., M., *Diffraction of Electrons by Perfect Crystals*, in *Electron Microscopy in MAterials Science II*, V. U. and R. E., Editors. 1975, CEC: Brussels. p. 397-552.

26. Bird, D.M. and Q.A. King, *Absorptive form factors for high-energy electron diffraction.* Acta Crystallographica Section A, 1990. **46**(3): p. 202-208.

27. Guo, Y., P.N.H. Nakashima, and J. Etheridge, *Three-beam convergent-beam electron diffraction for measuring crystallographic phases.* IUCrJ, 2018. **5**(6).

28. Moodie, A.F., J. Etheridge, and C.J. Humphreys, *The Symmetry of Three-Beam Scattering Equations: Inversion of Three-Beam Diffraction Patterns from Centrosymmetric Crystals.* Acta Crystallographica Section A, 1996. **52**(4): p. 596-605.

29. Zuo, J.M. and A.L. Weickenmeier, *On the Beam Selection and Convergence in the Bloch-Wave Method.* Ultramicroscopy, 1995. **57**(4): p. 375-383.

30. *Addtional information*. Available from: http://wrap.warwick.ac.uk/109737.

31. Gonzalez, R.C. and R.E. Woods, *Digital Image Processing*. 2008, Upper Saddle River, NJ 07458: Pearson Prentice Hall.

32. Pennington, R.S., et al., *Neural-network-based depth-resolved multiscale structural optimization using density functional theory and electron diffraction data.* Physical Review B, 2018. **97**(2): p. 024112.

33. Jansen, J., et al., *MSLS, a Least-Squares Procedure for Accurate Crystal Structure Refinement from Dynamical Electron Diffraction Patterns.* Acta Crystallographica Section A, 1998. **54**(1): p. 91-101.

34. Nüchter, W., A.L. Weickenmeier, and J. Mayer, *High-Precision Measurement of Temperature Factors for NiAl by Convergent-Beam Electron Diffraction.* Acta Crystallographica Section A, 1998. **54**(2): p. 147-157.

35. Shukla, R.C. and E. Sternin, *Debye-Waller factor in Cu: A Green's function approach.* Philosophical Magazine B, 1996. **74**(1): p. 1-11.

36. Martin, C.J. and D.A. O'Connor, *Anharmonic contributions to Bragg diffraction. I. Copper and aluminium.* Acta Crystallographica Section A, 1978. **34**(4): p. 500-505.

37. Spence, J.C.H. and J.M. Zuo, *Quantitative electron microdiffraction (Invited)*. Electron Microscopy 1994, Vol 1: Interdisciplinary Developments and Tools, ed. B. Jouffrey and C. Colliex. 1994. 837-838.

38. Saki, K., T. Kenji, and I. Nobuo, *Structural Evolution of Corundum at High Temperatures.* Japanese Journal of Applied Physics, 2008. **47**(1S): p. 616.

39. Barnea, Z.M.; Moss, G.R. , *Diffraction Studies of Real Atoms and Real Crystals*. 1974, Melbourne: Univ. of Melbourne.

40. Saravanan, R., S.K. Mohanlal, and K.S. Chandrasekaran, *Anharmonic Temperature Factors, Anomalous-Dispersion Effects and Bonding Charges in Gallium-Arsenide.* Acta Crystallographica Section A, 1992. **48**: p. 4-9.

41. Uno, R., T. Okano, and K. Yukino, *Electron Distribution in GaAs as Revealed by X-Ray Diffraction.* Journal of the Physical Society of Japan, 1970. **28**(2): p. 437-&.

42. Bublik, V.T. and S.S. Gorelik, *Über die röntgenographische Analyse der mittleren quadratischen Atomverschiebungen in einigen Halbleiterkristallen.* Kristall und Technik, 1977. **12**(8): p. 859-869.

43. Stahn, J., M. Mohle, and U. Pietsch, *Comparison of experimental and theoretical structure amplitudes and valence charge densities of GaAs.* Acta Crystallographica Section B-Structural Science, 1998. **54**: p. 231-239.

44. Saravanan, R., et al., *Electron density distribution in GaAs using MEM.* Journal of Physics and Chemistry of Solids, 2003. **64**(1): p. 51-58.

45. Schowalter, M., et al., *Computation and parametrization of the temperature dependence of Debye–Waller factors for group IV, III–V and II–VI semiconductors.* Acta Crystallographica Section A, 2009. **65**(1): p. 5-17.

46. Fazeli Kisomi, A. and S.J. Mousavi, *Ab initio calculations of the phonon and thermal properties of a (GaAs)1/(AlAs)1 superlattice and comparing them with the Ga0.5Al0.5As alloy.* Chinese Journal of Physics, 2017. **55**(3): p. 1062-1066.

47. Kushwaha, M.S., *Compressibilities, Debye-Waller factors, and melting criteria for II-VI and III-V compound semiconductors.* Physical Review B, 1981. **24**(4): p. 2115-2120.

48. Reid, J., *Debye-Waller factors of zinc-blende-structure materials - a lattice dynamical comparison.* Acta Crystallographica Section A, 1983. **39**(1): p. 1-13.

49. Schowalter, M., et al., *Computation and parametrization of the temperature dependence of Debye-Waller factors for group IV, III-V and II-VI semiconductors.* Acta Crystallographica Section A, 2009. **65**(1): p. 5-17.

50. Zuo, J.M. and J.L. Rouviere, *Solving difficult structures with electron diffraction.* Iucrj, 2015. **2**: p. 7-8.

51. Zuo, J.M., R. Hoier, and J.C.H. Spence, *3-Beam and Many-Beam Theory in Electron-Diffraction and its Use for Structure-Factor Phase Determination in Non-Centrosymmetric Crystal-Structures.* Acta Crystallographica Section A, 1989. **45**: p. 839-851.

52. Nakashima, P.N.H., A.F. Moodie, and J. Etheridge, *A practical guide to the measurement of structure phases and magnitudes by three-beam convergent beam electron diffraction.* Ultramicroscopy, 2008. **108**(9): p. 901-910.

53. Valset, K. and J. Tafto, *Bloch wave symmetries in electron diffraction: Applications to Friedels law, Gjonnes–Moodie lines and refraction at interfaces.* Ultramicroscopy, 2011. **111**(7): p. 854-859.

54. Spence, J.C.H., M. O'Keeffe, and J.M. Zuo, *Have orbitals really been observed?* Journal of Chemical Education, 2001. **78**(7): p. 877-877.

55. Capelli, S.C., et al., *Hirshfeld atom refinement.* IUCrJ, 2014. **1**(5): p. 361-379.